# Real-time myocardial landmark tracking for MRI-guided cardiac radio-ablation using Gaussian Processes

**Niek R.F. Huttinga[1,2], Osman Akdag[1], Martin F. Fast[1], Joost Verhoeff[1], Firdaus A. A. Mohamed Hoesein[3], Cornelis A.T. van den Berg[1,2], Alessandro Sbrizzi[1,2], Stefano Mandija*[1,2]**

[1]Department of Radiotherapy, University Medical Center Utrecht, Utrecht, The Netherlands.
[2]Computational Imaging Group for MR therapy & Diagnostics, University Medical Center Utrecht, Utrecht, The Netherlands.
[3]Department of Radiology, University Medical Center Utrecht, Utrecht, The Netherlands.
*Author to whom correspondence should be addressed. Email: s.mandija@umcutrecht.nl

## Abstract

*Objective.* The high speed of cardiorespiratory motion introduces a unique challenge for cardiac stereotactic radio-ablation (STAR) treatments with the MR-linac. Such treatments require tracking myocardial landmarks with a maximum latency of 100 ms, which includes the acquisition of the required data. The aim of this study is to present a new method that allows to track myocardial landmarks from few readouts of MRI data, thereby achieving a latency sufficient for STAR treatments. *Approach.* We present a tracking framework that requires only few readouts of k-space data as input, which can be acquired at least an order of magnitude faster than MR-images. Combined with the real-time tracking speed of a probabilistic machine learning framework called Gaussian Processes, this allows to track myocardial landmarks with a sufficiently low latency for cardiac STAR guidance, including both the acquisition of required data, and the tracking inference. *Main results.* The framework is demonstrated in 2D on a motion phantom, and in vivo on volunteers and a ventricular tachycardia (arrhythmia) patient. Moreover, the feasibility of an extension to 3D was demonstrated by in silico 3D experiments with a digital motion phantom. The framework was compared with template matching - a reference, image-based, method - and linear regression methods. Results indicate an order of magnitude lower total latency (<10 ms) for the proposed framework in comparison with alternative methods. The root-mean-square-distances and mean end-point-distance with the reference tracking method was less than 0.8 mm for all experiments, showing excellent (sub-voxel) agreement. *Significance.* The high accuracy in combination with a total latency of less than 10 ms – including data acquisition and processing - make the proposed method a suitable candidate for tracking during STAR treatments. Additionally, the probabilistic nature of the Gaussian Processes also gives access to real-time prediction uncertainties, which could prove useful for real-time quality assurance during treatments.

## 1. Introduction

Ventricular arrhythmia in cardiac patients is a major public health problem and a predominant cause of sudden cardiac death (Fishman *et al* 2010). These cardiac patients are treated with anti-arrhythmic medications, implantable cardioverter defibrillators (ICD), and undergo invasive catheter ablation procedures with the aim to ablate the arrhythmogenic substrate responsible for the arrhythmias (Cronin *et al* 2019). However, one major limitation for current catheter ablation techniques is the inaccessibility of the target area, for example at deep intramural or subepicardial locations in the heart (Tung *et al* 2015). Such target inaccessibility combined with the severe risks associated to invasive procedures currently make catheter ablations difficult or even impossible.

Recently, an alternative novel treatment option has been developed: non-invasive stereotactic arrhythmia radio-ablation (STAR) (Robinson *et al* 2019, Zei and Soltys 2017, Cuculich *et al* 2017). Such treatment consists of the delivery of high-dose radiotherapy (25 Gy) to the target area in a single fraction. First results showed a remarkable reduction of arrythmias after one year and limited acute toxicities (Robinson *et al* 2019). Currently, the safety and efficacy of this novel treatment technique is investigated from a clinical perspective within a multicentre consortium comprising more than 30 hospitals in Europe (STOPSTORM.eu), where around 100 patients have already been treated using this technique (Grehn *et al* 2022).

From a technical perspective, however, accurate radiation delivery to the target substrate during STAR treatments remains a major challenge due to cardiorespiratory motion, which results in an uncertainty on the target position. Currently, such uncertainties for the target position are partly accounted for by 1) managing respiratory motion via gated treatments, leading to long treatment times, and partly by 2) increasing the margins around the radiation target, leading to spilling of the prescribed dose to (possibly healthy) surrounding tissues (Lydiard *et al* 2021, Trojani *et al* 2022). To maximize dose delivery to the target and to minimize the toxicity to surrounding tissues, the following two key requirements are necessary: 1) target visualization for accurate dose planning, and 2) real-time cardiorespiratory motion characterization with low latency for treatment guidance.

In the context of STAR treatments, recent works have proposed to use the Cyberknife (Accuray, Inc., Sunnyvale, CA, USA) robotic CT-linac radiotherapy system and its Synchrony tracking software to track the tip of the lead of an implantable cardioverter-defibrillator (ICD). This tip is observable on CT images and acts as a surrogate of the motion of the non-visible target tissue (Dvorak *et al* 2022, Knybel *et al* 2021). From an imaging perspective, the superior soft-tissue contrast of MR systems compared to CT systems makes hybrid MR-linac systems another good candidate for STAR treatments, as they can allow accurate visualization of the target tissue (Puntmann *et al* 2016, Akdag *et al* 2022b), and thereby fulfil the first requirement posed above. A first case study already demonstrated that MR imaging highly improved the accuracy of the target definition, allowing beam-gating for treatment guidance and delivery (Mayinger *et al* 2020). However, this study also highlighted the difficulty of real-time treatment delivery based on direct tracking of the target due to cardiac motion and imaging artifacts arising from the implanted cardiac devices. Hence, especially of the second requirement for real-time guidance of STAR treatments, as posed above, is a major challenge for MR-based methods for STAR treatments.

Evidently, the most challenging aspect of real-time tracking methods is a low latency, i.e. a short time between the acquisition of the target location and the delivery of radiation to the target. The maximally acceptable latency naturally depends on the type of motion; the higher the motion speed and target displacement, the lower the acceptable latency. As of this writing, no commonly-accepted latency has been posed for cardiorespiratory motion management for STAR treatments yet. For respiratory motion, however, the AAPM Task Group 76 recommends a latency of 500 ms (Keall *et al* 2006). Since cardiac dynamics are approximately 4-5 times faster than respiratory dynamics, we deduce an acceptable latency of 100 ms for the real-time tracking of cardiorespiratory motion. To adhere to such latency, MR image acquisition, reconstruction and processing steps should be limited to 100 ms.

The current Unity MR-linac generation supports 2D cine MR imaging with temporal resolutions below 100 ms (Akdag *et al* 2021), and 3D cine MR imaging is currently not feasible at this speed. However, although the temporal image resolution is sufficient to resolve the cardiac motion, the latency of the multi-leaf collimator adjustment of the linac system is about 90 ms (Akdag *et al* 2022a). This thus effectively leaves only 10 ms latency for all other processing, including the MR image acquisition, reconstruction, and processing of the data for target tracking. Nowadays, several 2D and 3D image-based target tracking techniques have been proposed that rely on either rigid (Shi *et al* 2014, Ipsen *et al* 2016, Paganelli *et al* 2018) or deformable (Klein *et al* 2010, Zachiu *et al* 2015) image registration to estimate the motion of a predefined target. These methods, however, typically rely on images, and their total latency is therefore intrinsically affected by the MR imaging acquisition and reconstruction times. Consequently, it remains challenging to perform MR image acquisition and reconstruction with the desired latency of 10 ms, both in 2D and in 3D. Therefore, alternative approaches are needed, which ideally do not rely on MR images as input for target tracking, but rather use intermediate raw MR data.

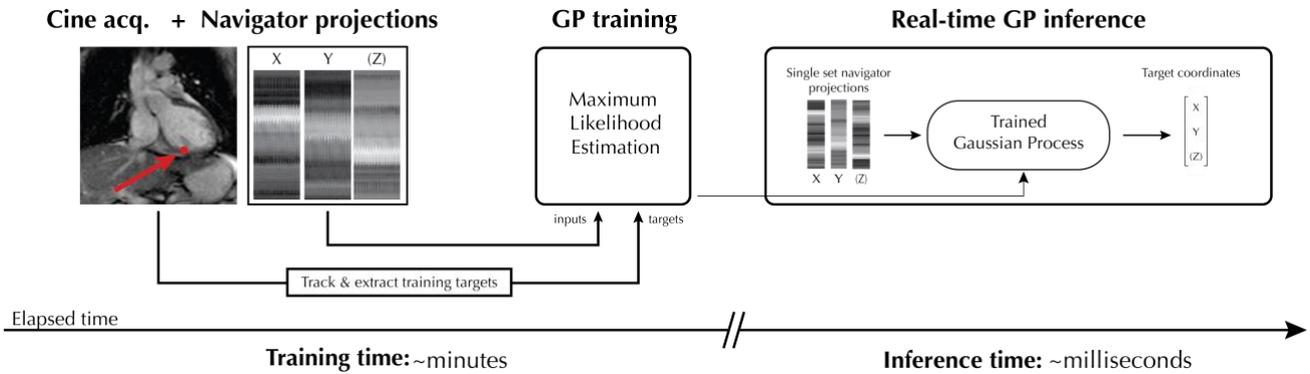

**Figure 1**: An overview of the framework. To construct the training set consisting of inputs and targets, first a cine acquisition is performed, interleaved with navigator projections. The inputs are defined as the raw data of the navigator projections. The targets (red arrow, bullet) are extracted using a reference tracking method, in this case template matching on the images. Next, the GP is trained on the training set using maximum likelihood estimation. Finally, during inference, the navigator projections are input to the trained GP, which then outputs the most-likely myocardial landmark and a corresponding measure of estimation uncertainty. The data required for inference can be acquired in 2/3x TR, and the GP inference takes less than a millisecond. This results in a total latency in the order of the repetition time.

The goal of this work is to demonstrate a proof-of-concept tracking method that requires as input only raw k-space data and could therefore achieve the required when combined with tracking inference. The method is based on a previous work, which investigated the application of a probabilistic machine learning methodology called Gaussian Processes (Rasmussen and Williams 2006) (GPs) to real-time respiratory motion estimation for radiotherapy. GPs have three properties that are highly desirable for real-time tracking: 1) real-time processing speeds in the order of milliseconds; 2) a real-time measure of estimation uncertainty; 3) a rapid training procedure using minimal data. The first property results from the completely analytic formulation of the inference procedure (Rasmussen and Williams 2006). The second property is inherent to GP's probabilistic nature. The GP inference is necessarily preceded by a training procedure. However, in contrast with deep learning methods, this training can be performed within minutes, on data that can be acquired in minutes. This third appealing property results from the relatively few parameters that a Gaussian Process requires.

The choice of input data is motivated by the fact that a single MRI k-space readout, which crosses the center of k-space, contains the information of the motion of all the excited tissues along that readout direction. Therefore, three mutually orthogonal readouts through the center of k-space, should contain information regarding the 3D motion of a target within the excited field of view (FOV). Initial results have demonstrated this feasibility and showed that such motion estimation could be performed in few milliseconds. The biggest benefit of such an approach is that it does not rely on the availability of a complete MR image and can rather deal with raw data that can be acquired at higher speeds (Huttinga *et al* 2022).

| Test | MR System | B0 | Acquisition type | Plane | Field of view [mm³] | Slices | Acquired voxel size [mm³] | TR / TE [ms / ms] | Flip angle [deg] | Pixel/Bandwidth [pix/hz] | Dynamics | Half scan factor | SENSE factor | Dynamic scan time [ms] |
|---|---|---|---|---|---|---|---|---|---|---|---|---|---|---|
| #1 | Unity MR-linac | 1.5 T | 2D Cartesian bFFE | Sagittal | 300 x 300 | 1 | 3 x 3 x 15 | 2.7 / 1.4 | 48 | 0.12 / 1894 | 1000 | 0.65 | 2.3 | 77 |
| #2 | Unity MR-linac | 1.5 T | 2D Cartesian bFFE | Coronal | 300 x 300 | 1 | 3 x 3 x 15 | 2.7 / 1.4 | 48 | 0.12 / 1894 | 1000 | 0.65 | 3 | 77 |
| #3 | Philips Elition | 3 T | 2D Cartesian bFFE | Coronal | 300 x 300 | 1 | 1.9 x 1.9 x 10 | 2.6 / 1.3 | 45 | 0.12 / 3420 | 400 | 0.65 | 2 | 110 |
| #4 | Unity MR-linac | 1.5 T | 2D Radial bFFE | Coronal | 350 x 350 | 1 | 3 x 3 x 10 | 3 / 1.5 | 50 | 0.12 / 1710 | 200 | no | 1 | 345 |

**Table 1**: An overview of the MR-acquisition parameters for all experiments described in the experiment section.

As mentioned previously, GPs not only allow for real-time tracking inference, but can also provide a real-time measure of the confidence of the estimated motion in real-time. Such a confidence measure could be of interest for quality assurance for STAR treatments (Huttinga *et al* 2022). Both properties make GPs exceptionally interesting for real-time radiotherapy applications such as the tracking of cardiac motion during STAR, as considered in this work.

In this work, we combine all desirable properties for real-time tracking methods described above into a single framework and propose to apply Gaussian Processes (GPs) to real-time myocardial landmark tracking. That is, tracking the spatial coordinates of predefined myocardial landmarks that ought to be tracked during STAR procedures, using data that can be acquired very rapidly. Since the framework requires only a few readouts of k-space data, and the inference is performed within milliseconds, it has the potential to perform myocardial landmark tracking at a framerate higher than 100 Hz, which is sufficient for cardiac radio-ablation. In this proof-of-concept, we demonstrate the proposed tracking method in 2D on a motion phantom and in-vivo (healthy volunteers and a VT patient). Finally, the feasibility of the extension to 3D landmark tracking with low latency is demonstrated in simulations through experiments with a digital 3D motion phantom.

## 2. Methods

### 2.1 Gaussian Processes for myocardial landmark tracking

We define the problem of tracking anatomical landmarks directly from k-space data and propose to solve it with GPs. This approach requires two steps: 1) patient-specific training, and 2) real-time inference. An overview of our approach is shown in Figure 1, and we describe it in more detail below.

First, given a training set consisting of landmark location coordinates $Y_D = \{y_D^i\}_{i=1}^{N_D}$, $y_D^i \subset \mathbb{R}^Y$, and fast k-space navigator readouts $X_D = \{x_D^i\}_{i=1}^{N_D}$, $x_D^i \subset \mathbb{R}^X$, the GP trains its internal parameters $\theta$ to approximate a function $f(x_D^i ; \theta) \approx y_D^i$, for all $i$. It does this by defining a Gaussian probability function over possible outputs using a so-called kernel function. This kernel function $K(x^i, x^j; \theta)$ takes two inputs $x^i, x^j \in \mathbb{R}^X$ and is parametrized by kernel hyper-parameters $\theta$. Effectively, the kernel function models the correlation between the inputs and the outputs. Many types of kernel functions are available, each resulting in different characteristics of the target function (e.g., smoothness, periodicity, etc.). In the training phase, the kernel's parameters are obtained by maximizing the likelihood of the resulting probability density function over the training targets $Y_D$ by tuning $\theta$ (Rasmussen and Williams 2006). In a subsequent inference phase, the trained GP is applied to evaluate the posterior probability distribution over the landmark locations, given the training data and the newly acquired, unseen, k-space readouts $X_Q = \{x_Q^i\}_{i=1}^{N_Q}$, $x_Q^i \subset \mathbb{R}^X$. In other words, the GP estimates the most-likely landmark location, $f(x_Q^i; \theta)$, and a corresponding measure of uncertainty, defined as the width of the Gaussian posterior probability density around that estimate. Intuitively, the output landmark locations are computed as a weighted average over all landmark locations in the training set, where the weights are determined by the learned kernel function; the more similar the training samples are to the newly acquired data according to the learned kernel function, the higher the weights in the weighted average computation. For more details on the inner workings of the Gaussian Process we refer to (Huttinga *et al* 2022, Rasmussen and Williams 2006).

### 2.2 Gaussian Processes training, testing, and evaluation

Application of GPs requires a paired training set $D$ which includes inputs (k-space readouts) and "ground-truth" outputs (myocardial landmarks), i.e. $D := \{(x_D^i, y_D^i)\}_{i=1}^{N_D}$. The landmarks $y_D^i$ are obtained from a reference tracking method. In this work, template matching (TM) was used as ground-truth for training and evaluation of the GP (see below) (Shi *et al* 2014). TM used normalized cross-correlation in a 3x3 cm² search region around the landmark of interest (magenta marked target in the figures) to automatically track a

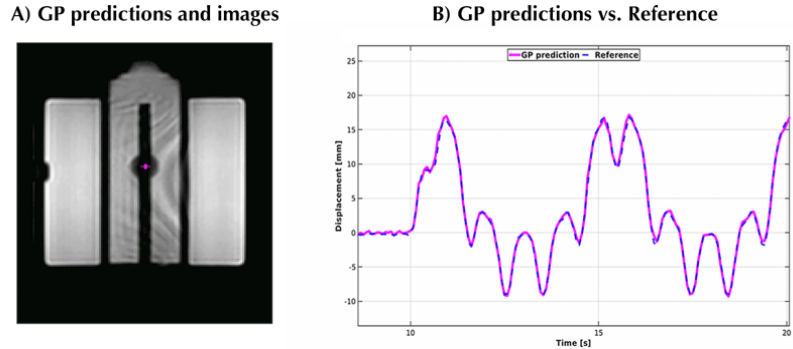

**A) GP predictions and images**   **B) GP predictions vs. Reference**

**Figure 2:** 2D landmark (magenta point in part a) tracking using a motion phantom on the MR-linac. GP prediction from two orthogonal, retrospectively simulated spokes and comparison with TM. This figure is a snapshot of a GIF, and we recommend viewing this animated figure here as Animated Figure 1.

predefined target. Prior to TM, images were median filtered (3x3) and upsampled (x2).

The GP inputs are defined as two mutually orthogonal spokes in 2D, and three mutually orthogonal spokes in 3D, either retrospectively simulated or prospectively acquired. Independent GPs were trained for each one of the landmark coordinates, that is, two independent GPs for 2D coordinates, three independent GPs for 3D coordinates. It should be noted, however, that also the correlation between directions can be modelled with multi-output GPs (Bonilla *et al* 2007). However, we have empirically observed that this did not improve results and only made the processing steps more complex. In all experiments, a Matérn 3/2 kernel was used with automatic relevance determination (ARD). This type of kernel function allows to model reasonably smooth underlying processes (at least once mean-square differentiable (Rasmussen and Williams 2006)), and the ARD allows to automatically determine the k-space components along the input spokes that are most informative (relevant) for the output prediction.

For each experiment described below, cine MR images were acquired, and corresponding location targets and GP inputs were extracted as described above. Next, each acquired dataset was split into two contiguous, disjoint, equally-sized sets: the train and the test set, respectively. The GPs were trained on the first half: the training set. The training time per GP was about 35 s for experiments 1 and 2, resulting in a total of about 70 s training time per experiment. For experiment 3, the training time was 10 s per GP, while for experiment 4 the training time was 5 s per GP. Please see below for more details regarding the experiments.

After training, testing was performed on the second half of the data. Given the complete analytic formulation of the GP inference procedure(Rasmussen and Williams 2006), this inference step can be performed within 0.1 ms per dynamic. All processing was performed in Matlab, and for all GP processing, the optimized implementations of the GPML Matlab toolbox by Rasmussen and Nickisch (2010) were used.

To validate the presented method, the GP output was compared with TM on the hold-out test set, used here as a reference. For each dynamic, the root-mean-square-distance (RMSD) and the end-point-distance (EPD) (Euclidean distance between vectors) metrics were computed between the positions obtained with these two methods.

## 3. Experiments

Six different experiments were performed as described below. Details about the adopted MR sequences are reported in Table 1. The first four experiments focus on the application of GP in 2D. In these experiments, we build up in complexity: first we started from a phantom experiment and retrospective landmark tracking; then we tested the GP framework on

**A) GP predictions and images**    **B) GP vs. reference feet-head**    **C) GP vs. reference right-left**

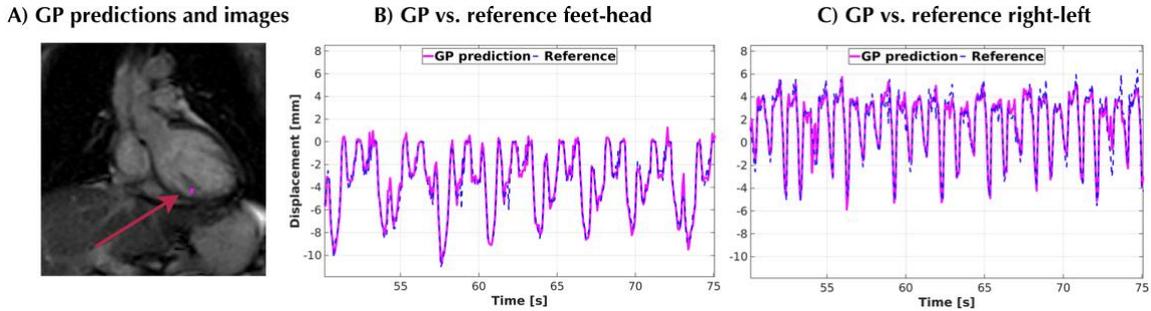

**Figure 3:** in vivo 2D landmark tracking on a healthy volunteer on the MR-linac. GP prediction (magenta line) of a landmark in the myocardium (magenta point in part a, indicated with the red arrow) from two orthogonal, retrospectively simulated spokes for a healthy volunteer. A comparison is made with TM (reference, blue line) for the FH (b) and LR (c) directions. This figure is a snapshot of a GIF, and we recommend viewing this animated figure here as Animated Figure 2.

retrospective volunteer data; after we tested the GP framework on a retrospective cardiac patient dataset; next, we tested the GP framework for 2D prospective landmark tracking. In the fifth experiment, we demonstrate the capability of the proposed methodology to cope with 3D data in a controlled setting (simulations). This is presented to show the potential of the methodology. Finally in the sixth experiment we test the capability of GP to provide uncertainty of the predicted landmark positions for a 2D case.

## 3.1 Retrospective real-time target tracking

### 3.1.1 Real-time tracking of a motion phantom

In this first experiment, the feasibility for GP-based myocardial landmark tracking was retrospectively validated with 2D Cartesian cine-MR images acquired on a 1.5 T Unity MR-linac system (Elekta AB, Stockholm, SE) using an MR-compatible 4D motion phantom (ModusQA, Ontario, Canada). The phantom contained a spherical insert, used here as target, and was programmed to move according to a (ground-truth) waveform defined as the sum of a $\cos^4$ waveform mimicking cardiac motion (60 bpm, 10 mm peak-to-peak amplitude), and a $\cos^4$ waveform mimicking respiratory motion (12 bpm, 20 mm peak-to-peak amplitude) (Akdag *et al* 2021). A GP was trained to track a myocardial landmark (magenta-marked in Figure 2), from two retrospectively simulated k-space spokes along FH and LR. Template matching was performed with a circular Hough transform (Davies 2004) on the motion phantom's insert to train and validate the GP.

### 3.1.2 Retrospectively undersampled data, healthy volunteer

In the second experiment, in vivo retrospective myocardial landmark tracking was demonstrated in a healthy volunteer from data acquired on an MR-linac. This experiment served to test the feasibility of in vivo tracking. A GP was trained to track the magenta-marked myocardial target from two retrospectively simulated k-space spokes along FH and LR. Target locations obtained from TM were used for training the GPs.

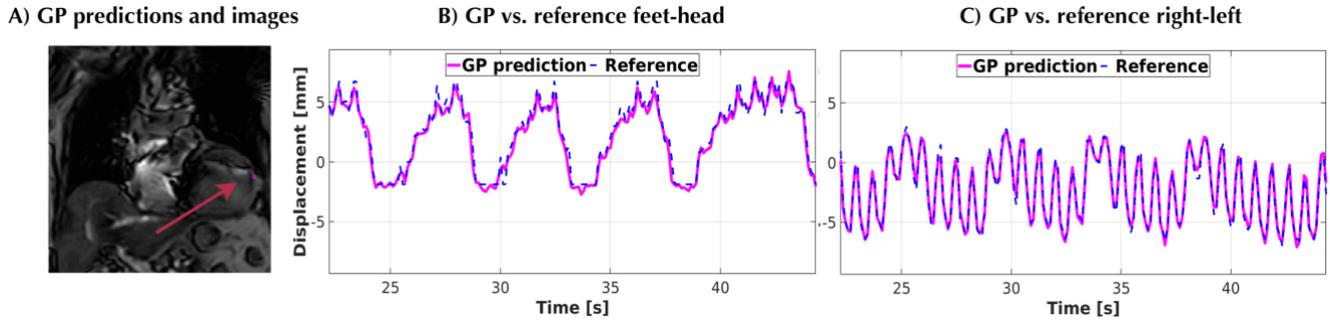

**Figure 4:** in vivo 2D landmark tracking on a cardiac patient on a diagnostic MR system. GP prediction (magenta line) of a landmark in the myocardium (magenta point in part a, indicated with the red arrow) from two orthogonal, retrospectively simulated spokes for the cardiac patient. A comparison is made with TM (reference, blue line) for the FH (b) and LR (c) directions. This figure is a snapshot of a GIF, and we recommend viewing this animated figure here as Animated Figure 3.

### 3.1.3 Retrospectively undersampled data, VT patient

In the third experiment, in vivo retrospective tracking was demonstrated on a patient with cardiac arrhythmia, who had received cardiac radio-ablation. This experiment served to test the robustness of GP against cardiac implant-related artifacts. The patient had a cardiac valve, an implanted cardiac device and related leads. A GP was trained to track the magenta-marked myocardial target from two retrospectively simulated k-space spokes along FH and LR. Target locations obtained from TM were used for training the GPs.

### 3.2 Prospective real-time target tracking

In the fourth experiment, prospective in vivo 2D myocardial landmark tracking was demonstrated in a healthy volunteer using an MR-linac. This experiment served to test the feasibility of prospective in vivo tracking, as opposed to the retrospective experiments described above. A 2D uniform radial cine-MRI was acquired, and GP inputs were extracted from raw MR-data. A single channel of raw k-space data of two prospectively acquired spokes along FH and LR were selected as input. Target locations obtained from TM were used for training the GPs.

### 3.3 Other experiments and evaluation

### 3.3.1 In silico 3D tracking

In the fifth experiment, the feasibility of an extension to 3D landmark tracking was investigated in silico. This experiment served to test the feasibility of an extension of the presented framework to real-time 3D tracking, as this would be ideal for guidance during STAR. To this end, the XCAT digital phantom (Segars *et al* 2010) was used to simulate 3D cardiorespiratory motion. The following parameters were used for this simulation: voxel size: 3x3x3 mm³, temporal resolution 10 ms, cardiac motion: 80 bpm, 30 mm peak-to-peak amplitude; respiratory motion parameters were set as: 12 bpm, 60 mm peak-to-peak amplitude. Data were simulated along FH, AP, LR spokes by computing projections over the image, and a GP was trained to track the magenta-marked myocardial target from the three spokes. Note that only for this case, the GP was trained and validated on ground-truth landmark locations and not on locations resulting from the TM step; in other words, the training targets for this simulation are noiseless.

### 3.3.2 Comparison against other regression methods

For the 2D in vivo experiments described above, the presented GP regression-based method was compared against two alternatives: linear regression, and linear regression preceded by a principal component analysis (PCA) compression. For the linear regression, a linear model was fit on the training set to predict the landmark coordinates using the same navigator spokes as the GP as input. More specifically, a matrix $X$ is constructed by placing navigator inputs $x_b^i$ from the training set - as defined in Section 2.1 - in its $i$-th row, and a matrix $Y$ is constructed by placing landmark targets $y_b^i$ from the training set - as also defined in Section 2.1 - in its $i$-th row. The linear model's coefficients $c$ were obtained with the pseudo-inverse $X^+$ of $X$:

$$c = X^+ \cdot Y.$$ Additionally, a linear model was fit to the landmarks in the training set by using a compressed representation of the input matrix $X$. To this extent, $X$ was compressed in the column direction, prior to performing the fit. To compress the matrix, the columns were projected onto their principal components in the column direction, thereby varying the number of components over all possible values. This effectively reduced the number of parameters of the model to the number of selected principal components, potentially improving its generalization from training to test set. Finally, this truncated linear model was fit to the same targets $Y$ using pseudo-inverse of the compressed matrix, similarly as above. For both linear regression methods, the inference on a newly-observed set of navigator spokes $x_Q$ was performed as $y_Q = x_Q \cdot c$, for model coefficients $c$.

For both alternative methods, the performance was evaluated in the same manner as the GP: for each dynamic, the root-mean-square-distance (RMSD) and the end-point-distance (EPD) (Euclidean distance between vectors) metrics were computed between their respective outputs and the reference method template matching.

### 3.3.3 Uncertainty estimation during bulk motion

In the sixth experiment, the effect of a bulk motion event on the GP's prediction uncertainty was investigated, to see whether this could be used as a measure of quality assurance. To this extent, we used the healthy volunteer data of experiment 2, and used the first 40 s for training, and the data from the remaining 40 s for testing. The training set comprised normal cardiorespiratory motion, while in the testing set bulk motion was simulated by gradually shifting the image 10 voxels in the LR direction (17 mm, see Figure 8). Since the model was trained on normal cardiorespiratory motion, we hypothesise that this bulk motion shift will increase the GP's uncertainty of landmark location estimates, demonstrating how the GP's uncertainty could be used for quality assurance.

## 4. Results

This section summarizes the results of the experiments. For all considered experiments, the training step took less than a minute, and the required training data was acquired in less than a minute too. The inference step, including data acquisition and processing, took less than 10 ms.

### 4.1 Retrospective real-time target tracking

#### 4.1.1 Real-time tracking of motion phantom

The phantom results in Figure 2 on two retrospectively computed orthogonal k-space spokes indicate the feasibility of prospective landmark tracking within 10 ms with GPs (acquisition time: 2 x TR = 5.4 ms; GP prediction time: <1 ms). The GP predictions show low RMSD (FH direction: 0.24 mm) and EPD (0.28±0.15 mm) when compared to the independently computed landmarks with TM, used as a reference.

#### 4.1.2 Retrospectively undersampled data, healthy volunteer

Figure 3 shows the performance of in vivo GP myocardial landmark tracking from two orthogonal k-space spokes retrospectively computed from fully sampled k-space data on a healthy volunteer and its comparison with TM as independent reference. The presented results show the feasibility of GP to accurately track a selected landmark (magenta point in the figure). As previously observed, the in vivo GP-based landmark predictions are obtained within 10 ms and show small, sub-voxel, differences with TM: RMSD-FH=0.36 mm, RMSD-LR=0.44 mm, and EPD=0.40±0.24 mm.

#### 4.1.3 Retrospectively undersampled data, VT patient

Figure 4 shows the performance of GP myocardial landmark tracking from two orthogonal k-space spokes retrospectively computed from fully sampled k-space data on a cardiac patient and its comparison with TM used as independent reference. The presented results show the feasibility of GP to accurately track a selected landmark (magenta point in the figure) and demonstrate the robustness of GP to image artifacts arising from the implanted cardiac device. As for the previous experiment, GP-based landmark predictions are obtained within 10 ms and the computed values for the distance metrics show low RMSD (FH direction = 0.42 mm, LR direction = 0.39 mm) and EPD (0.33±0.20 mm).

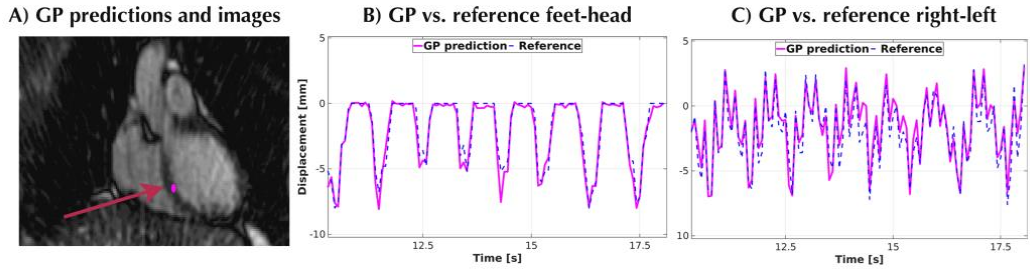

**Figure 5:** In vivo 2D landmark tracking on prospectively acquired healthy volunteer data on the MR-linac. GP prediction (magenta line) of a landmark in the myocardium (magenta point in part a, indicated with the red arrow) from two orthogonal spokes extracted from the uniform radial acquisition. A comparison is made with TM (reference, blue line) for the FH (b) and LR (c) directions. This figure is a snapshot of a GIF, and we recommend viewing this animated figure here as Animated Figure 4.

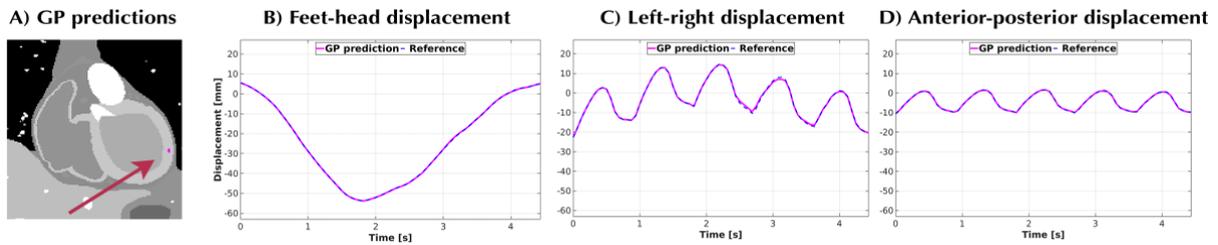

**Figure 6:** In silico feasibility test of 3D landmark tracking. GP predictions (magenta lines) of a landmark in the myocardium (magenta point in part a, indicated with the red arrow) from three orthogonal spokes are compared to the ground truth target location (reference, blue line) for the FH (b), LR (c) and AP (d) directions. This figure is a snapshot of a GIF, and we recommend viewing this animated figure here as Animated Figure 5.

## 4.2 Prospective real-time target tracking

Figure 5 shows the feasibility of prospective GP-based myocardial landmark tracking using two orthogonal k-space spokes acquired using a uniform radial acquisition on a healthy volunteer. The good agreement between GP and TM used as independent reference (RMSD-FH=0.47 mm; RMSD-LR=0.76 mm; EPD=0.69±0.39 mm) demonstrates the capability of GP to perform prospective myocardial landmark tracking with similar accuracy as image-based TM, while allowing tracking times within the desired latency of 10 ms (acquisition: 2x TR = 6 ms; GP prediction: <1ms).

## 4.3 Other experiments and evaluation

### 4.3.1 In silico 3D tracking

Figure 6 shows in silico the feasibility of 3D GP-based myocardial landmark tracking using three orthogonal spokes computed along the FH, AP, LR directions using the XCAT digital phantom. These results demonstrate sub-voxel agreement with the ground truth target location (RMSD-FH=0.21 mm; RMSD-LR=0.37 mm; RMSD-AP=0.18 mm; EPD=0.19±0.10 mm).

### 4.3.2 Comparison against other regression methods

The comparison against the two other regression methods, linear regression, and linear regression+pca, is shown in Figure 7. The GP-based regression seems to perform as good or better than the best-case scenario of the other methods. However, besides the good accuracy, the GP also provides a measure of uncertainty quantification. This could be useful for quality assurance during treatments, and is not naturally available for the other methods.

### 4.3.3 Uncertainty estimation during bulk motion

In Figure 8, the effect of bulk motion on GP uncertainty is evaluated through simulation using the data from experiment 2 (healthy volunteer). This figure shows that when an unforeseen event like bulk motion occurs, the computed uncertainty increases (>15% increase). This can therefore be used as a prospective measure of quality assurance of the computed target location with GPs.

### 4.3.4 End-point-error overview

In Figure 9, an overview of the end-point-distances and RMSD for the above experiments is provided. This shows that for each experiment, the observed distances are smaller than the image voxel size, indicating the excellent agreement between the GP's and TM's target location predictions.

## 5. Discussion and Conclusions

A successful clinical workflow for STAR requires real-time cardiorespiratory target tracking and radiation delivery (<100 ms). Given that the latency of the multi-leaf collimator adjustment of the Elekta Unity MR-linac system is about 90 ms (Akdag *et al* 2022a), one of the remaining technical challenges is to develop an imaging/tracking technique with only 10 ms latency, including the acquisition of the required data. It should be noted that most known tracking methods require image inputs, which – even in 2D – cannot be acquired and reconstructed within 10 ms with MRI.

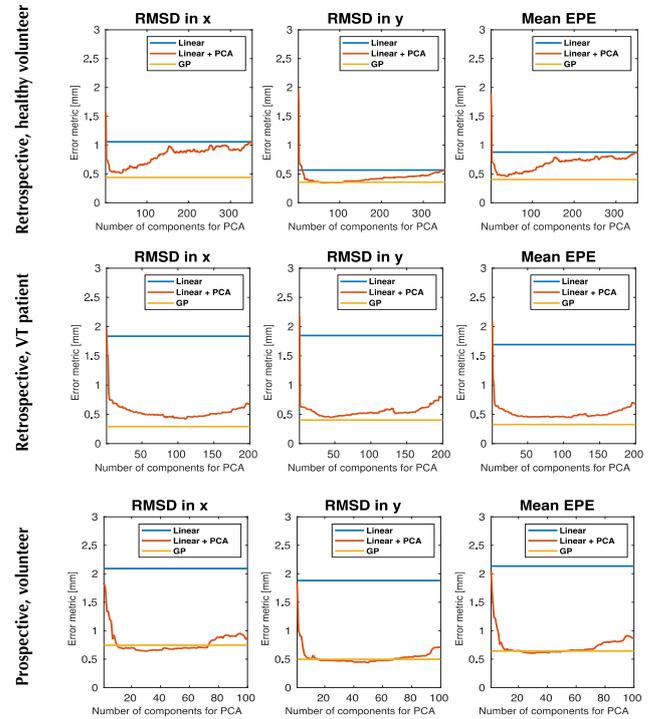

**Figure 7:** Comparison of GP-based regression (yellow) with linear regression (blue) and linear regression preceded by PCA (linear + pca, red). The comparison is performed for three different experiments (rows) and evaluated using RMSD in x (left-right) and y (feet-head), and EPE (columns).

In this work, we introduced a proof-of-concept tracking method with less than 10 ms total latency (acquisition and processing), based on a probabilistic machine learning method called Gaussian Processes. The method only requires raw k-space data as input and is thereby less bound to the imaging speed of MRI. We have demonstrated the technical feasibility of 2D landmark tracking with 10 ms total latency and sub-voxel differences with conventional image-based methods. The method was evaluated with a phantom, healthy volunteers, and a VT patient. It should be noted that the VT patient's images had degraded quality due to the ICD and leads. Nevertheless, the method showed a robust performance even in this case, thereby further indicating the potential for cardiac radio-ablation on VT patients. Ultimately, we have also shown the feasibility of prospective 2D myocardial landmark tracking on a 1.5T Elekta Unity MR-linac for guidance of STAR treatments. Finally, in silico experiments indicate the feasibility of an extension to 3D myocardial landmark tracking. However, it should be mentioned that the experiments in this work are only preliminary. Future work should focus on the demonstration of prospective, real-time in vivo 3D myocardial landmark tracking on volunteers.

The proposed framework requires a patient-specific training and an inference phase. We envision that the required patient-specific training data may be acquired during the conventional day-to-day treatment verification, prior to any radiotherapy treatment. It should be noted that GPs are data-efficient, and only minimal training data is required that can usually be acquired in less than a minute (see Table 1). Subsequently, the template matching can be performed automatically in about half a minute, and then GP training can be performed within a minute. This would result in a training phase that can be performed within 5 minutes, in parallel to the verification of the treatment delivery. This is in stark contrast with deep learning-based methods, that frequently require large databases and days or weeks of training. During the treatment, the GP's inference frequency of landmark locations is mostly governed by the acquisition frequency of navigator spokes required for inference. Hence, the navigator spokes may be interleaved with conventional image acquisitions to facilitate both GP-based landmark location inference at the desired temporal frequency, and conventional image acquisitions for visual aid.

A still relatively unexploited output of the presented framework is the uncertainty quantification. Preliminary results in previous work (Huttinga *et al* 2022), indicate that the uncertainty estimates may be used for online quality assurance, e.g. by temporarily halting the treatment in case of high uncertainty and resuming it whenever confidence is restored. This is also evident from the bulk motion experiment in this work; the uncertainty increased significantly during the bulk motion. This property may be interesting for STAR treatments on VT patients, where VT storms may occur during treatment, but also in general to detect unexpected motion events that could affect the treatment quality directly. This topic will be the subject of future work.

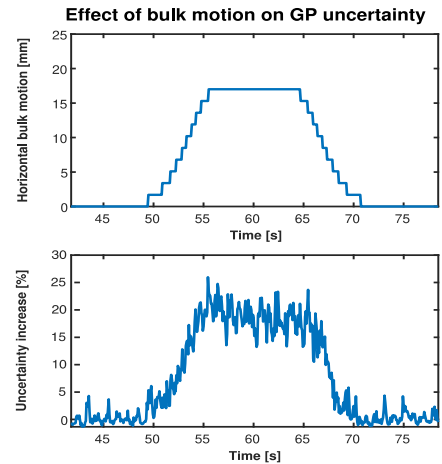

**Figure 8**: The effect of bulk motion on GP uncertainty evaluated through simulation using the data from experiment 2 (healthy volunteer). Top graph: simulation of bulk motion as image shift in the LR direction. Bottom graph: percentual uncertainty increase with respect to the mean uncertainty during training.

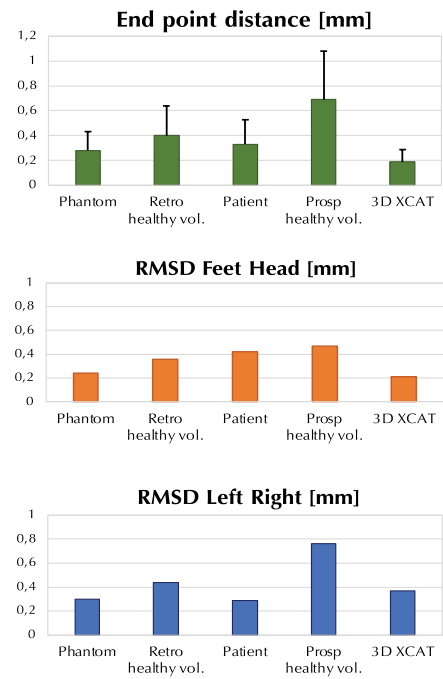

**Figure 9**: Overview of the end-point-distance for the previous five experiments (top) and RMSD FH

Several other points should also be addressed in future work. The proposed method shows potential for real-time target tracking due to fast processing times, and the minimal amount of required data which can be acquired within few milliseconds. However, to truly achieve low latency during inference, the required navigator spokes should be acquired at the desired high temporal frequency, rather than interleaved with conventional imaging readouts, as is currently done. Preliminary results have indicated that this is feasible for 3D respiratory motion at 67 Hz (Huttinga *et al* 2022), but a complete practical low-latency implementation should nevertheless be demonstrated in a future work on this specific application.

An unavoidable consequence of using a training set to calibrate a model, is that the model's output quality (i.e. error w.r.t. ground-truth location) will be similar to that of the training data. Gaussian Process have the favourable ability to filter out Gaussian noise on the targets (Rasmussen and Williams 2006), but especially non-Gaussian deviations of the training targets from the ground-truth targets – e.g. due to structurally wrong target locations, or several large jumps in the target locations – may result in a loss of performance. We have empirically observed that the template matching quality may indeed be improved in several scenarios, which could eventually also improve the GP's performance. One obvious alternative to TM would be manual tracking. Albeit labour-intensive and undesirable in practice, this could give more insight on the relation between training data quality and final accuracy.

A related matter that should be addressed in future work is the extension to an in vivo 3D setting. The main challenge there is training data, since this would require 3D images of the heart at sufficient temporal resolution to resolve cardiorespiratory motion. Fortunately, previous work indicates the feasibility of training on images of average breathing, and estimating time-resolved dynamics (Huttinga *et al* 2022). A similar strategy could therefore be employed for the extension to real-time 3D myocardial landmark tracking.

## Acknowledgements

The authors acknowledge funding by ITEA Eureka cluster on Software innovation through the SIGNET project no. 20052, and by the Dutch Research Council (NWO) through project no. 18078 (VENI) and project no. 19484 (MEGAHERTZ).

## Ethical statement

The volunteers provided written informed consent prior to all scans, all scans were approved by the institutional review board of the University Medical Center Utrecht, and were carried out in accordance with the relevant guidelines and regulations.

## References

Akdag O, Borman P T S, Woodhead P, Uijtewaal P, Mandija S, Asselen B V, Verhoeff J J C, Raaymakers B W and Fast M F 2022a First experimental exploration of real-time cardiorespiratory motion management for future stereotactic arrhythmia radioablation treatments on the MR-linac *Phys. Med. Biol.* **67** 065003

Akdag O, Mandija S, Borman P T, Alberts E and Fast M F 2021 Feasibility of free breathing real-time cine-MRI for MR-guided cardiac radioablation on the Unity MR-linac *Proc. Intl. Soc. Mag. Reson. Med* vol 29 p 4014

Akdag O, Mandija S, van Lier A L H M W, Borman P T S, Schakel T, Alberts E, van der Heide O, Hassink R J, Verhoeff J J C, Mohamed Hoesein F A A, Raaymakers B W and Fast M F 2022b Feasibility of cardiac-synchronized quantitative T1 and T2 mapping on a hybrid 1.5 Tesla magnetic resonance imaging and linear accelerator system *Phys. Imaging Radiat. Oncol.* **21** 153–9

Bonilla E V, Chai K and Williams C 2007 Multi-task Gaussian process prediction *Adv. Neural Inf. Process. Syst.* **20**

Cronin E M, Bogun F M, Maury P, Peichl P, Chen M, Namboodiri N, Aguinaga L, Leite L R, Al-Khatib S M, Anter E, Berruezo A, Callans D J, Chung M K, Cuculich P, d'Avila A, Deal B J, Della Bella P, Deneke T, Dickfeld T-M, Hadid C, Haqqani H M, Kay G N, Latchamsetty R, Marchlinski F, Miller J M, Nogami A, Patel A R, Pathak R K, Sáenz Morales L C, Santangeli P, Sapp J L, Sarkozy A, Soejima K, Stevenson W G, Tedrow U B, Tzou W S, Varma N, Zeppenfeld K, ESC Scientific Document Group, Asirvatham S J, Sternick E B, Chyou J, Ernst S, Fenelon G, Gerstenfeld E P, Hindricks G, Inoue K, Kim J J, Krishnan K, Kuck K-H, Avalos M O, Paul T, Scanavacca M I, Tung R, Voss J, Yamada T and Yamane T 2019 2019 HRS/EHRA/APHRS/LAHRS expert consensus statement on catheter ablation of ventricular arrhythmias *EP Eur.* **21** 1143–4

Cuculich P S, Schill M R, Kashani R, Mutic S, Lang A, Cooper D, Faddis M, Gleva M, Noheria A and Smith T W 2017 Noninvasive cardiac radiation for ablation of ventricular tachycardia *N. Engl. J. Med.* **377** 2325–36

Davies E R 2004 *Machine Vision: Theory, Algorithms, Practicalities* (Elsevier)

Dvorak P, Knybel L, Dudas D, Benyskova P and Cvek J 2022 Stereotactic Ablative Radiotherapy of Ventricular Tachycardia Using Tracking: Optimized Target Definition Workflow *Front. Cardiovasc. Med.* **9** 870127

Fishman G I, Chugh S S, DiMarco J P, Albert C M, Anderson M E, Bonow R O, Buxton A E, Chen P-S, Estes M, Jouven X, Kwong R, Lathrop D A, Mascette A M, Nerbonne J M, O'Rourke B, Page R L, Roden D M, Rosenbaum D S, Sotoodehnia N, Trayanova N A and Zheng Z-J 2010 Sudden Cardiac Death Prediction and Prevention: Report From a National Heart, Lung, and Blood Institute and Heart Rhythm Society Workshop *Circulation* **122** 2335–48

Grehn M, Mandija S, Andratschke N, Zeppenfeld K, Blamek S, Fast M, Botrugno C, Blanck O, Verhoeff J and Pruvot E 2022 Survey results of the STOPSTORM consortium about stereotactic arrhythmia radioablation in Europe *EP Eur.* **24** euac053.376

Huttinga N R F, Bruijnen T, van den Berg C A T and Sbrizzi A 2022 Gaussian Processes for real-time 3D motion and uncertainty estimation during MR-guided radiotherapy *ArXiv Prepr. ArXiv220409873*

Ipsen S, Blanck O, Lowther N J, Liney G P, Rai R, Bode F, Dunst J, Schweikard A and Keall P J 2016 Towards real-time MRI-guided 3D localization of deforming targets for non-invasive cardiac radiosurgery *Phys. Med. Biol.* **61** 7848

Keall P J, Mageras G S, Balter J M, Emery R S, Forster K M, Jiang S B, Kapatoes J M, Low D A, Murphy M J and Murray B R 2006 The management of respiratory motion in radiation oncology report of AAPM Task Group 76 a *Med. Phys.* **33** 3874–900

Klein S, Staring M, Murphy K, Viergever M A and Pluim J P W 2010 elastix: A Toolbox for Intensity-Based Medical Image Registration *IEEE Trans. Med. Imaging* **29** 196–205

Knybel L, Cvek J, Neuwirth R, Jiravsky O, Hecko J, Penhaker M, Sramko M and Kautzner J 2021 Real-time measurement of ICD lead motion during stereotactic body radiotherapy of ventricular tachycardia *Rep. Pract. Oncol. Radiother.* **26** 128–37

Lydiard S, Pontré B, Lowe B S, Ball H, Sasso G and Keall P 2021 Cardiac radioablation for atrial fibrillation: Target motion characterization and treatment delivery considerations *Med. Phys.* **48** 931–41

Mayinger M, Kovacs B, Tanadini-Lang S, Ehrbar S, Wilke L, Chamberlain M, Moreira A, Weitkamp N, Brunckhorst C, Duru F, Steffel J, Breitenstein A, Alkadhi H, Garcia Schueler H I, Manka R, Ruschitzka F, Guckenberger M, Saguner A M and Andratschke N 2020 First magnetic resonance imaging-guided cardiac radioablation of sustained ventricular tachycardia *Radiother. Oncol.* **152** 203–7

Paganelli C, Lee D, Kipritidis J, Whelan B, Greer P B, Baroni G, Riboldi M and Keall P 2018 Feasibility study on 3D image reconstruction from 2D orthogonal cine-MRI for MRI-guided radiotherapy *J. Med. Imaging Radiat. Oncol.* **62** 389–400

Puntmann V O, Peker E, Chandrashekhar Y and Nagel E 2016 T1 mapping in characterizing myocardial disease: a comprehensive review *Circ. Res.* **119** 277–99

Rasmussen C E and Nickisch H 2010 Gaussian Processes for Machine Learning (GPML) Toolbox *J. Mach. Learn. Res.* **11** 3011–5

Rasmussen C E and Williams C K I 2006 *Gaussian Processes for Machine Learning* vol 2 (Cambridge, MA: MIT Press) Online: https://direct.mit.edu/books/book/2320/Gaussian-Processes-for-Machine-Learning

Robinson C G, Samson P P, Moore K M S, Hugo G D, Knutson N, Mutic S, Goddu S M, Lang A, Cooper D H, Faddis M, Noheria A, Smith T W, Woodard P K, Gropler R J, Hallahan D E, Rudy Y and Cuculich P S 2019 Phase I/II Trial of Electrophysiology-Guided Noninvasive Cardiac Radioablation for Ventricular Tachycardia *Circulation* **139** 313–21

Segars W P, Sturgeon G, Mendonca S, Grimes J and Tsui B M W 2010 4D XCAT phantom for multimodality imaging research *Med. Phys.* **37** 4902–15

Shi X, Diwanji T, Mooney K E, Lin J, Feigenberg S, D'Souza W D and Mistry N N 2014 Evaluation of template matching for tumor motion management with cine-MR images in lung cancer patients *Med. Phys.* **41** 052304

Trojani V, Botti A, Grehn M, Balgobind B, Savini A, Pruvot E, Verhoeff J, Iori M and Blanck O 2022 Stereotactic arrhythmia radioablation in europe: treatment planning benchmark results of the STOPSTORM consortium *EP Eur.* **24** euac053.371

Tung R, Vaseghi M, Frankel D S, Vergara P, Di Biase L, Nagashima K, Yu R, Vangala S, Tseng C-H, Choi E-K, Khurshid S, Patel M, Mathuria N, Nakahara S, Tzou W S, Sauer W H, Vakil K, Tedrow U, Burkhardt J D, Tholakanahalli V N, Saliaris A, Dickfeld T, Weiss J P, Bunch T J, Reddy M, Kanmanthareddy A, Callans D J, Lakkireddy D, Natale A, Marchlinski F, Stevenson W G, Della Bella P and Shivkumar K 2015 Freedom from recurrent ventricular tachycardia after catheter ablation is associated with improved survival in patients with structural heart disease: An International VT Ablation Center Collaborative Group study *Heart Rhythm* **12** 1997–2007

Zachiu C, Papadakis N, Ries M, Moonen C and Senneville B D de 2015 An improved optical flow tracking technique for real-time MR-guided beam therapies in moving organs *Phys. Med. Biol.* **60** 9003

Zei P C and Soltys S 2017 Ablative Radiotherapy as a Noninvasive Alternative to Catheter Ablation for Cardiac Arrhythmias *Curr. Cardiol. Rep.* **19** 79